\documentstyle[latexsym,amssymb,aps,floats,eqsecnum,
preprint,amsfonts]{revtex}
\def\laq{\raise 0.4ex\hbox{$<$}\kern -0.8em\lower 0.62 ex\hbox{$\sim$}}
\def\gaq{\raise 0.4ex\hbox{$>$}\kern -0.7em\lower 0.62 ex\hbox{$\sim$}}
\input epsf
\begin{document}

\bibliographystyle{unsrt}

\title{Vector field localization and negative tension branes}
\author{Massimo Giovannini
\footnote{Electronic address: 
Massimo.Giovannini@ipt.unil.ch}}

\address{{\it Institute of Theoretical Physics, 
University of Lausanne}}
\address{{\it BSP-1015 Dorigny, Lausanne, Switzerland}}

\maketitle

\begin{abstract} 
It is shown that negative tension branes in higher 
dimensions may lead to an effective lower 
dimensional theory where the gauge-invariant vector fields 
associated with the fluctuations of the metric
are always massless and localized on the brane. 
Explicit five-dimensional examples of this phenomenon are 
provided. Furthermore, it is 
shown that higher dimensional gauge fields 
can also be localized on these configurations 
with the zero mode separated from the massive tower by a gap. 
\end{abstract}
\vskip0.5pc
\centerline{{\bf To appear in Physical Review D}}
\vskip0.5pc
\noindent

\newpage
\renewcommand{\theequation}{1.\arabic{equation}}
\setcounter{equation}{0}
\section{Introduction}
Suppose that the space-time contains non-compact 
extra-dimensions and consider, for sake of 
concreteness the following $D$-dimensional geometry
\begin{equation}
ds^2 = a^2(w)[dt^2 - dx_{1}^2 -...- dx_{d}^2 - dw^2],
\label{metric}
\end{equation}
with one bulk coordinate $w$ and with $D= d +2$ 
space-time dimensions. 
Even in the absence of gravitational interactions, 
five-dimensional domain-wall solutions allow 
the localization of fermionic zero modes \cite{m1}. 
This type of multi-dimensional domain-wall 
solutions can be obtained by setting the warp factors 
to a constant. From a strictly gravitational point 
of view the four-dimensional Planck mass would not be finite.
In this sense the increase of the dimensionality 
of space-time represents a tool in order to 
obtain an effective lower dimensional theory 
where chiral fermions may be successfully localized.
Indeed, this approach led to a very interesting 
development in the analysis of chiral gauge theories on the lattice.
In particular, it was shown that the massless fermionic 
zero mode is still present if the five-dimensional 
continuous space is replaced by a lattice \cite{kap}. Hence, it
was realized that chiral symmetry can be realized in different ways avoiding 
the known problem of doubling of fermionic degrees of freedom \cite{nin}.
A crucial ingredient, in this context, was the use of the so-called
Wilson-Ginzparg relation \cite{wg}. 

In spite of the fact that an increase in the dimensionality 
of space-time may lead to localized chiral fermions a problem may still 
remain as far as vector fields are concerned \cite{rubrev}. In particular, 
a higher-dimensional domain-wall solution of the type of \cite{m1} 
does not lead to the localization of vector fields. More specifically, neither 
the vector fields generated as fluctuations of the higher-dimensional metric, 
nor the  gauge fields can be localized on the same wall to which the
fermionic zero-modes are attached. The localization of vectors fields 
is the subject of the present analysis. It is 
well known that there are various mechanisms aiming at the localization 
of gauge fields in higher-dimensional contexts \cite{dv,od,dub,dg,kt}. 
In most of these approaches the idea is to analyze the situation
where the four-dimensional Planck mass
\begin{equation}
M_{P}^2 \simeq M^3 \int d w a^{d}(w), 
\label{plm}
\end{equation}
is finite. In such a  background it can be shown, in general terms, that
the graviton zero mode is also localized  leading to ordinary 
gravity on the wall \cite{rs1,rs2}. A particularly simple example 
of this phenomenon is provided by five-dimensional AdS space where, however, 
neither the gauge fluctuations coming from 
 the geometry \cite{mg,mg2,mg3} nor the minimally 
coupled higher dimensional gauge fields can be simply localized \footnote{ The 
reason for this statement is that the squared modulus of the zero mode
leads, in this case, to a logarithmically divergent normalization integral.}.

Suppose now to drop this hypothesis. Suppose, in other words, that 
the four-dimensional Planck mass is not finite. Is it possible 
to localize spin one fields on such a configuration? As in 
\cite{m1,kap} gravitational interactions are 
 used here as an effective tool in order to 
localize higher spins but {\em not} as an interaction which 
should be, by itself, localized.
In this sense, the present approach is 
more modest than the one proposed in \cite{dv,od,dub,dg,kt} where 
{\em all } the higher spins should be, in principle, localized.

Our logic is, in short, the following. Consider, for sake of concreteness, 
a five-dimensional warped geometry of the type of Eq. (\ref{metric}) with 
$d =3$. In this type of warped geometry
vector modes may come both from the 
 gauge field living in the bulk and from  the vector fields 
arising as fluctuations of the five-dimensional metric.
In qualitative terms, the possible behavior of spin one 
fluctuations of the geometry and of bulk gauge fields will be now  separately 
examined.

The 
spin one fluctuations of  a five-dimensional geometry in the presence 
of a brane configuration can be expressed in terms of two 
equations which are written in fully 
gauge-invariant [i.e. coordinate independent] 
terms\footnote{The Greek indices label the $(3 +1)$ (Poincar\'e invariant)
dimensions. The Latin (uppercase) indices run over the whole 
five-dimensional space-time.}:
\begin{eqnarray}
&&\partial_{\alpha} \partial^{\alpha} {\cal V}_{\mu} =0,
\nonumber\\
&&{\cal V}_{\mu}' + 
\frac{3}{2} {\cal H} {\cal V}_{\mu}=0,
\label{calV}
\end{eqnarray}
where the prime denotes the derivative with respect 
to $w$ and ${\cal H} = (\ln{a})'$. Eq. (\ref{calV}) will be 
specifically derived in Section II and generalized in Section III. 
Here some qualitative remarks will be made.

Eqs. (\ref{calV}) tell that the vector modes of the geometry 
are always massless and that the zero mode
 is localized provided 
\begin{equation}
\int_{0}^{\infty} \frac{d w}{a^3(w)}, 
\label{norm}
\end{equation}
is convergent. By looking, simultaneously, at Eq. (\ref{norm}) and 
at the expressions of the curvature invariants pertaining to the geometry 
(\ref{metric})
\begin{eqnarray}
&& R = \frac{4}{a^2} ( 2 {\cal H}' + 3 {\cal H}^2),
\nonumber\\
&& R_{M N}R^{M N} = \frac{4}{a^4}( 9 {\cal H}^4 + 6 {\cal H}' {\cal H}^2
+ 5 {{\cal H}'}^2),
\nonumber\\
&& R_{M N A B}R^{M N A B} = \frac{8}{a^4}( 
2 {{\cal H}'}^2 - 5 {\cal H}^4),
\label{curv}
\end{eqnarray}
it can be argued that there may exist warped backgrounds where the 
vector modes of the geometry are localized and the bulk geometry is regular
\footnote{By regular we mean that the curvature invariant are analytic 
functions of the bulk coordinate for every $w>0$. For $w=0$ 
we will allow the possibility of a singularity which comes from the thin
nature of the brane configuration.}.
An example of this class of backgrounds is the warp factor 
\begin{equation}
a(w) = e^{K |w|},
\label{warp}
\end{equation}
with $K>0$. One of the questions which will be 
investigated in this paper is the actual possibility of realizing such 
a model. As it will be shown in Section II this type of 
behavior will never be realized in original Randall-Sundrum set-up.
This statement will be made clear by a detailed calculation. Here it 
is enough to notice that in the Randall-Sundrum model the $Z_{2}$ symmetry, 
together with the requirement that the four-dimensional Planck mass is finite,
implies that while the integral appearing in Eq. (\ref{plm}) is always 
{\em convergent}, the integral of Eq. (\ref{norm}) is always 
{\em divergent} (in particular for $w\to \infty$). Hence, the graviphoton 
is never normalizable.

The gauge fields leaving in the bulk lead to spin one 
fluctuations obeying equations which are different from the 
ones of the spin one fluctuations arising directly from the geometry.
In order to have an explicit example, consider,
for instance, the case  of an Abelian gauge field coupled to the geometry 
in five dimensions. The action of such a field can be written as 
\begin{equation}
S = - \frac{1}{4} \int d^{5}x \sqrt{ - G} F_{A B}F^{A B},
\label{gaugeac}
\end{equation}
where $ F_{A B} = \nabla_{[A} A_{B]}  \equiv \partial_{[A} A_{B]}$ and $A_{B}$ 
is the vector potential. From Eq. (\ref{gaugeac}),  
a Schr\"odinger-like  equation leading can be obtained for the mass 
eigenstates. 
Defining $\Delta_{1} 
= \sqrt{a}$,  the normal modes of the action (\ref{gaugeac}) are 
${\cal A}_{\mu} = \Delta_{1}
 A_{\mu}$. In terms of these normal modes the equation of motion is 
\begin{equation}
{\cal A}_{\mu}'' - \frac{\Delta_1''}{\Delta_1} {\cal A}_{\mu} -
 \partial_{\alpha} \partial^{\alpha} {\cal A}_{\mu} =0,
\label{geq1}
\end{equation}
where $\partial_{\mu} {\cal A}^{\mu} =0 $ and where ${\cal A}_{w} =0$.
In the specific case of the action
 (\ref{gaugeac}) $ \Delta_1 = \sqrt{a}$ but, in more  general terms, 
some couplings to the higher-dimensional dilaton field could be 
expected. This case will be studied in Section IV. Here only notice
that the normalization condition of Eq. (\ref{geq1})
for the gauge zero mode leads to the following integral
\begin{equation}
\int_{0}^{\infty} d w \Delta_1^2. 
\label{int}
\end{equation}
In the case of five-dimensional AdS space-time (i.e. 
$a(w) \simeq 1/|w|$, $\Delta_1^2 \sim a~ \sim 1/|w|$) the integral (\ref{int})
 is logarithmically divergent. Is it possible to envisage the situation 
where $\Delta \sim e^{ -K_1 |w|}$? In this case 
the gauge zero mode would be  localized and it would be 
separated from the massive tower by a mass gap of the order $K_1$. 
The problem of the gauge zero modes and the possible occurrence 
of a mass gap in the spectrum will be one of the issues related to the 
present analysis.

A perspective similar to the one discussed in this 
paper has been recently adopted in a partially related 
problem, namely the possibility of implementing 
the Higgs mechanism without invoking a fundamental scalar \cite{mp}. 
Suppose \cite{mp}
that the higher dimensional theory gives rise to a weight function 
multiplying the gauge kinetic term by a factor which depends upon
the bulk coordinate $w$. In the simplest realization this weight function 
can be related to the volume of the extra-dimensional space but there may 
be different situations like the ones where some other field 
(like the dilaton) gives important contributions. If the weight 
function satisfies a number of properties, then, the pure higher dimensional 
theory may lead to a lower-dimensional theory of massive vector bosons
with a gap in the spectrum.  Another important analogy is that 
the backgrounds leading to the effects analyzed in \cite{mp} do not 
necessarily lead to a finite four-dimensional Planck mass. 

The plan of the  paper is the following. In Section II  
arguments will be provided in order to show that the gauge
modes of the geometry may be localized {\em provided} the four-dimensional 
Planck mass is not finite. In Section III an explicit model 
of negative tension branes will be provided. In Section IV the localization 
of the spin one fluctuations  
will be discussed. Section V contains some concluding
remarks.

\renewcommand{\theequation}{2.\arabic{equation}}
\setcounter{equation}{0}
\section{Basic Considerations}

Consider, as a warm-up, the case of gauge-invariant vector fluctuations 
in the case where the bulk action contains the usual Einstein-Hilbert 
term supplemented by a bulk cosmological constant in the presence 
of a thin brane configuration. This is exactly 
the case of a Randall-Sundrum type of background and it will be shown that, 
in this case the vector fluctuations of the geometry cannot be localized, as 
anticipated in the introduction.

The action for  the system is 
given as 
\begin{equation}
S = \int d^{5} x \sqrt{|G|}\biggl[ - R - \Lambda\biggr] 
- \frac{\lambda}{2} \int 
d^4 \rho \sqrt{|\gamma|}[\gamma^{\alpha\beta} \partial_{\alpha} X^{M}
\partial_{\beta} X^{N} G_{M N}-2], 
\end{equation}
where $\gamma_{\alpha\beta}$ is the induced metric on the brane 
and where $\rho^{\mu}$ are the coordinates on the brane. It is important 
to write down the action in rigorous terms not so much for the analysis 
of the background but for the analysis of the vector fluctuations. 
By functional derivation of the action with respect to the space-time 
and induced metric we get a set of two (coupled) equations that can be 
simply written as
\begin{eqnarray}
&&{\cal Q}_{A B} = {\cal T}_{A B},
\label{ein}\\
&& \gamma_{\alpha\beta} = \partial_{\alpha}X^{M} \partial_{\beta} X^{N} 
G_{M N}
\label{ind}
\end{eqnarray}
where 
\begin{eqnarray}
&&{\cal Q}_{A B} = R_{A B}
 + \frac{\Lambda}{3} G_{A B} 
\label{qdef}\\
&& {\cal T}_{A B} = \frac{\lambda}{2} \int d^4 \rho 
\frac{\sqrt{|\gamma|}}{\sqrt{|G|}}\gamma^{\alpha\beta} \partial_{\alpha}X^{M}
\partial{\beta}X^{N}( G_{A M} G_{B N} - \frac{1}{3}G_{M N} G_{A B})\delta(w).
\label{tdef}
\end{eqnarray}
In the background  (\ref{metric}) with $d=3$ the induced metric 
is  
\begin{equation}
\gamma_{\alpha\beta} = \delta^{A}_{\alpha} 
\delta^{B}_{\beta} \eta_{A B} a(w)^2,
\end{equation}
and  the equations for the background are
\begin{eqnarray}
&& 6 ( {\cal H}^2 - {\cal H}') = \lambda a \delta(w),
\label{b1}\\
&& 6[ 3 {\cal H}^2 + {\cal H}'] + 2 \Lambda a^2 + \lambda a \delta(w) =0.
\label{b2}
\end{eqnarray}
In the case of five-dimensional AdS space-time, the only solution to Eqs. 
(\ref{b1})--(\ref{b2}) is given by 
\begin{equation}
a(w) = \frac{1}{( b |w | + 1)}
\label{rsback}
\end{equation}
where  $b = \sqrt{- \Lambda/12}$. 

Consider now the localization properties of the 
vector fluctuations pertaining to  this geometry. 
In order to address this question Eqs. (\ref{ein}) 
and (\ref{ind})  have to be perturbed 
with respect to the spin one degrees of freedom appearing 
in the fluctuations of $ G_{A B}$, namely,
\begin{equation}
\delta G^{(V)}_{A B}= a^2(w) \biggl(\matrix{ 
2\partial_{(\mu} f_{\nu)} 
& D_{\mu}  &\cr
D_{\mu}  & 0 &\cr} \biggr).
\label{pertv}
\end{equation}
where $\partial_{(\mu} f_{\nu)} = \frac{1}{2}(\partial_{\mu} f_{\nu} +
\partial_{\nu} f_{\mu}) $ and where $\partial_{\alpha} f^{\alpha} =0$ and 
$\partial_{\alpha} D^{\alpha} =0$. Eq. (\ref{pertv}) represents 
the most general decomposition of the spin one fluctuations of the geometry. 
The remaining nine degrees of freedom correspond either to scalar or to
tensor fluctuations. 

From Eqs. (\ref{ein})--(\ref{ind}) the perturbed equations are:
\begin{eqnarray}
&&\delta {\cal Q}_{A B} = \delta {\cal T}_{A B},
\label{p1}\\
&& \delta \gamma_{\mu\nu} = \delta^{M}_{\mu} \delta^{N}_{\nu} \delta G_{M N}. 
\label{p2}
\end{eqnarray}
As far as the vector fluctuations are concerned, the relevant components 
are the $(\mu,w)$ and the $(\mu, \nu)$.
Direct calculations show that
\begin{eqnarray}
&& \delta {\cal Q}_{\mu\nu} = \biggl[ \frac{2}{3} \Lambda a^2 
+ 2 ( {\cal H}' + 3 {\cal H}^2) \biggr]
\partial_{(\mu}f_{\nu)} + [ \partial_{(\mu} f_{\nu)}']'  
+ 3 {\cal H} [ \partial_{(\mu} f_{\nu)}']
\nonumber\\
&& -  [ \partial_{(\mu} D_{\nu)}]'  - 3 {\cal H} [ \partial_{(\mu} D_{\nu)}],
\label{q1}\\
&& \delta {\cal Q}_{ \mu w} = [ \frac{\Lambda}{3} a^2  + {\cal H}' + 
3 {\cal H}^2] D_{\mu}- 
\frac{1}{2} \partial_{\alpha} \partial^{\alpha} [ D_{\mu} - f_{\mu}'],
\label{q2}
\end{eqnarray}
and that 
\begin{eqnarray}
&& \delta {\cal T}_{\mu\nu} = - \frac{\lambda}{3} a(w) \delta(w)
 \partial_{(\mu}f_{\nu)},
\label{t1}\\
&& \delta {\cal T}_{\mu w} = - \frac{\lambda}{6} a(w) \delta(w) D_{\mu}. 
\end{eqnarray}
Hence, the relevant equations for the spin one fluctuations of the metric
 can be written as 
\begin{eqnarray}
&& -\frac{1}{2} \partial_{\alpha}\partial^{\alpha} [ D_{\mu} - f_{\mu}'] 
+ [{\cal H}' + 3 {\cal H}^2 + \frac{\Lambda}{3} a^2 + \frac{\lambda}{6} 
a \delta(w) ] D_{\mu}=0,
\label{p1a}\\
&& -  [ \partial_{(\mu} D_{\nu)}]'  - 3 {\cal H} [ \partial_{(\mu} D_{\nu)}]
+[ \partial_{(\mu} f_{\nu)}']'  + 3 {\cal H} [ \partial_{(\mu} f_{\nu)}']
\nonumber\\
&&+ \biggl[ \frac{2}{3} \Lambda a^2 + 2 ( {\cal H}' + 3 {\cal H}^2) + 
\frac{\lambda}{6} a \delta(w)\biggr]
\partial_{(\mu}f_{\nu)} =0.
\label{p2a}
\end{eqnarray}
The vectors $f_{\mu}$ and $ D_{\mu}$ are not invariant with 
respect to infinitesimal coordinate transformations preserving the vector
 nature of the fluctuations. The coordinate transformation preserving 
the vector nature of the fluctuations is a nothing but a vector shift  
in the coordinates which can be parametrized in terms of the 
 (divergence-less) Poincar\'e vector $\zeta_{\mu}$:
\begin{equation}
x_{\mu} \to \tilde{x}_{\mu} = x_{\mu} + \zeta_{\mu}. 
\label{tra}
\end{equation}
On the fixed background defined by Eqs. (\ref{b1})--(\ref{b2}), Eq. 
(\ref{tra}) induces a computable transformation in $f_{\mu}$ and $D_{\mu}$ 
while we move from the original coordinate system to the 
tilded one. More specifically, the shift is given by 
\begin{eqnarray}
&&\tilde{f}_{\mu} = f_{\mu} - \zeta_{\mu},
\label{g1}\\
&& \tilde{D}_{\mu} = D_{\mu} - \zeta_{\mu}'.
\label{g2}
\end{eqnarray}
Now looking together at Eqs. (\ref{g1})--(\ref{g2}) we can define 
a gauge-invariant vector fluctuation which is 
exactly
\begin{equation}
{\cal V}_{\mu} = a^{3/2}[ D_{\mu} - f_{\mu}'].
\label{ginv}
\end{equation}
This quantity is clearly invariant under infinitesimal 
coordinate transformations. 
The factor $a^{3/2}$ in Eq. (\ref{ginv}) comes about since we want to 
work with fluctuations which have canonical kinetic terms in the action 
\footnote{
Since the 
kinetic term of the 
combination $D_{\mu} - f_{\mu}'$ always appears, in the action, 
multiplied by $a^3(w)$, 
${\cal V}_{\mu}$ is correctly normalized.}.
In terms of the gauge-invariant fluctuations Eqs. (\ref{p1}) and (\ref{p2}) 
can be written in a fully coordinate independent form  since the  
gauge-dependent part vanishes once the background equations 
(\ref{b1})--(\ref{b2}) 
are used inside the perturbed equations of motion (\ref{p1a})--(\ref{p2a}). 
The result is that 
\begin{eqnarray}
&& \partial_{\alpha} \partial^{\alpha} {\cal V}_{\mu} =0,
\nonumber\\
&& {\cal V}_{\mu}' + \frac{3}{2} {\cal H} {\cal V}_{\mu} =0.
\end{eqnarray}
Hence, as anticipated in the introduction, the gauge-invariant 
vector fluctuations are always massless and the 
normalization condition for the zero mode implies that 
\begin{equation}
\int_{0}^{\infty} \frac{ d  w}{a^3(w)}, 
\label{rsnorm}
\end{equation}
should be finite. In the case of the set-up 
discussed in the present Section, inserting Eq. (\ref{rsback}) into 
(\ref{rsnorm}), the resulting integral is never convergent.
Moreover, the form of Eq. (\ref{b1})--(\ref{b2}) prevents a solution 
of the type $a(w) \simeq e^{K |w|}$ for the warp factor. 

As discussed in the introduction, the evolution equation for the 
canonical gauge modes corresponding to the action of an
Abelian field strength can be simply obtained
\begin{equation}
{\cal A}_{\mu}'' - \frac{(\sqrt{a})''}{\sqrt{a}} {\cal A}_{\mu} -
 \partial_{\alpha} \partial^{\alpha} {\cal A}_{\mu} =0,
\end{equation}
and the normalization condition for the zero mode 
implies that the integral 
\begin{equation}
\int_{0}^{\infty} a(w) d w 
\end{equation}
should be convergent. Again, using Eq. (\ref{rsback}) it can be checked that 
the above integral is always logarithmically divergent. 
The conclusion is that, in the case of the typical Randall-Sundrum set-up 
neither the vector fields of the geometry nor the bulk 
gauge fields can be normalized. 
 
\renewcommand{\theequation}{3.\arabic{equation}}
\setcounter{equation}{0}
\section{Negative tension branes}
Consider now a slightly different situation, namely the 
case where the brane action also contain a dilatonic 
contribution. 
The action of the system, in analogy with the notations of the 
previous Section can be written as:
\begin{eqnarray}
S &=& \int d^{5} x \sqrt{|G|} \biggl[ - R + \frac{1}{2} G^{ A B} 
\partial_{A}\varphi
\partial_{B} \varphi - V(\varphi) \biggr]
\nonumber\\
&-& \frac{\lambda}{2} f(\varphi) \int d^{4} \rho \sqrt{|\gamma|} [
\gamma^{\alpha\beta} \partial_{\alpha} X^{M}  \partial_{\beta} X^{N} 
G_{M N} - 2],
\end{eqnarray}
whose related equations of motion can be written as 
\begin{eqnarray}
&& {\cal Q}_{A B} = {\cal T}_{A B} ,
\label{nb1}\\
&& G^{ A B} ( \partial_{A} \partial_{B} \varphi - \Gamma_{A B }^{C} 
\partial_{C} \varphi) +
\frac{\partial V}{\partial\varphi} + 
\frac{\lambda}{2} \frac{\partial f}{\partial\varphi} 
 \int d^{4} \rho \frac{\sqrt{|\gamma|}}{\sqrt{|G|}}
 \gamma^{\alpha\beta} \partial_{\alpha}X^{M} 
\partial_{\beta} X^{N} G_{M N} \delta(w) =0,
\label{nb2}\\
&& \gamma_{\alpha\beta} = f(\varphi) \partial_{\alpha} X^{A} 
\partial_{\beta} X^{B} G_{A B},
\label{nb3} 
\end{eqnarray}
where, now, 
\begin{eqnarray}
{\cal Q}_{A B } &=& R_{A B} - \frac{1}{2} \partial_{A} 
\varphi\partial_{B}\varphi 
+ \frac{V}{3} G_{A B}, 
\label{defq}\\
{\cal T}_{ A B} &=&  \frac{\lambda}{2} f(\varphi) \int d^4 \rho 
\frac{ \sqrt{|\gamma|}}{\sqrt{|G|} }
\gamma^{\alpha\beta} \partial_{\alpha} X^{M} \partial_{\beta} X^{N} 
[ G_{A N} G_{B M} 
- \frac{1}{3} G_{M N} G_{A B} ]\delta(w).
\label{deft}
\end{eqnarray}
Consider now the case where the dilaton field only depends upon the 
bulk coordinate. Then Eqs.  (\ref{nb1})--(\ref{nb2}) give 
\begin{eqnarray}
&& {\varphi'}^2 = 6 ( {\cal H}^2 - {\cal H}') - 
\lambda a(w) f^2(\varphi) \delta(w) ,
\label{nb1a}\\
&& {\cal H}' + 3 {\cal H}^2 + \frac{V}{3} a^2 + \frac{\lambda}{6} 
a f^2(\varphi) \delta(w) =0,
\label{nb2a}\\
&& \varphi'' + 3 {\cal H}\varphi' - \frac{\partial V}{\partial\varphi}a^2 - 
2 \lambda \frac{\partial f}{\partial \varphi} a \delta(w) =0,
\label{nb3a}
\end{eqnarray} 
with
\begin{equation}
\gamma_{\alpha\beta} = a^2(w) f(\varphi) \delta^{M}_{\alpha} 
\delta^{N}_{\beta} \eta_{ A B}.
\label{nind}
\end{equation}
With the same algebra discussed in the previous Section we can obtain the
 equation 
describing the gauge-invariant vector fluctuations. Indeed, recalling that 
\begin{eqnarray}
&& \delta {\cal Q}_{\mu\nu} = \biggl[ \frac{2}{3} V a^2 + 2 ( {\cal H}' 
+ 3 {\cal H}^2) \biggr]
\partial_{(\mu}f_{\nu)} + [ \partial_{(\mu} f_{\nu)}']' 
 + 3 {\cal H} [ \partial_{(\mu} f_{\nu)}']
\nonumber\\
&& -  [ \partial_{(\mu} D_{\nu)}]'  - 3 {\cal H} [ \partial_{(\mu} D_{\nu)}],
\label{nq1}\\
&& \delta {\cal Q}_{ \mu w} = [ \frac{V}{3} a^2  + {\cal H}' + 3 {\cal H}^2] - 
\frac{1}{2} \partial_{\alpha} \partial^{\alpha} [ D_{\mu} - f_{\mu}'],
\label{nq2}
\end{eqnarray}
and that 
\begin{eqnarray}
&& \delta {\cal T}_{\mu\nu} = - \frac{\lambda}{3} a(w)f^2(\varphi)
 \delta(w) \partial_{(\mu}f_{\nu)},
\label{nt1}\\
&& \delta {\cal T}_{\mu w} = - \frac{\lambda}{6} a(w) f^2(\varphi)
 \delta(w) D_{\mu}, 
\label{nt2}
\end{eqnarray}
exactly the same equations for the gauge-invariant vector fluctuations
 can be obtained.

The analysis of the background equations leads in the case examined in the 
present Section to the possibility of the solutions postulated  in the 
introduction on the basis of more general arguments.
More specifically, consider the following parameterization 
of the solution to  Eq. (\ref{nb1a})--(\ref{nb3a}):
\begin{eqnarray}
&&a(w) = e^{ B | w|},~~~~~~ \varphi(w) = A | w|,
\label{ana}\\
&& V(\varphi) = \Lambda e^{ b\varphi},~~~~~~ f(\varphi) = e^{ c \varphi}
\label{anf}
\end{eqnarray}
Inserting Eqs. (\ref{ana})--(\ref{anf}) into Eqs. (\ref{nb1a})-(\ref{nb3a})
we obtain the following matching conditions for the 
discontinuities :
\begin{eqnarray}
&& B = - \frac{\lambda}{12},
\label{c1}\\
&& A = c \lambda,
\label{c2}
\end{eqnarray}
and the following algebraic equations from the remaining terms:
\begin{eqnarray}
&& A^2 = 6 B^2,~~~~~ B^2 = - \frac{\Lambda}{9} ,
\label{c3}\\
&& 2 B + b A =0, ~~~~~ 3 A B - b \Lambda =0.
\label{c4}
\end{eqnarray}
The algebraic relations (\ref{c3}) and (\ref{c4}) to gather with 
the junction condition (\ref{c1})--(\ref{c2}) 
give rise to two sets of solutions depending upon the sign of $A$. If 
$A>0$ we will have increasing dilaton solutions (denoted by $A^{>}$). 
If $A < 0 $ we will have decreasing dilaton solutions (denoted $A^{<}$).

Let us now consider, separately, these two classes of solutions.
In the first case the solution of Eqs. (\ref{c1})--(\ref{c2}) together 
with Eqs. (\ref{c3})--(\ref{c4}) can be written as 
\begin{eqnarray}
&& A^{>} = \sqrt{\frac{2}{3}} \sqrt{ - \Lambda}, 
\label{a+}\\
&& c^{>} = \mp \frac{1}{4} \sqrt{\frac{2}{3}} \equiv\frac{1}{4} b^{>},
\label{c+}\\
&& B^{>}_{\pm} = \pm \sqrt{ - \frac{\Lambda}{9}}, 
\label{B+}\\
&& \lambda^{>}_{\pm} = \mp 4 \sqrt{-\Lambda},
\label{l+}
\end{eqnarray}
If $A<0$ the corresponding solution reads 
\begin{eqnarray}
&& A^{<} = -\sqrt{\frac{2}{3}} \sqrt{ - \Lambda}, 
\label{a-}\\
&& c^{<} = \pm \frac{1}{4} \sqrt{\frac{2}{3}} \equiv\frac{1}{4} b^{<},
\label{c-}\\
&& B^{<}_{\pm} = \pm \sqrt{ - \frac{\Lambda}{9}}, 
\label{B-}\\
&& \lambda^{<}_{\pm} = \mp 4 \sqrt{-\Lambda}.
\label{l-}
\end{eqnarray}
Notice that the relation among the couplings  $c$ and $b$ [i.e. $c = b/4$]
holds in both sets of solutions. In Eqs. (\ref{a+})--(\ref{l+}) 
and also in Eqs. (\ref{a-})--(\ref{l-})  the $\pm$ refers to the 
sign of $B$. Hence, for both sets of solutions $B$  is 
{\em positive} for {\em negative tension branes} [i.e. $\lambda < 0$ ] 
and it is 
{\em negative} for {\em positive tension branes} [i. e. $ \lambda > 0$]. 
Notice that our conventions differ slightly from the ones usually 
employed in the case of positive tension dilatonic walls \cite{cpl}.

\renewcommand{\theequation}{4.\arabic{equation}}
\setcounter{equation}{0}
\section{Localization properties of spin one fluctuations}
Let us now analyze the case of negative tension branes in light 
of the localization properties of the vector fields of the geometry. 
If the brane has negative tension the gauge-invariant vector zero mode 
is always localized. In fact, since the zero mode is 
\begin{equation}
{\cal V}_{\mu}^{0} = \frac{{\cal N}_{1} }{ a(w)^{3/2}},
\end{equation}
the corresponding normalization condition
\begin{equation}
{\cal N}_{1}^2\int_{-\infty}^{\infty} \frac{ d w}{a^3(w)} 
= \frac{ 2}{3 B_{+}} \equiv 
\frac{2}{\sqrt{ - \Lambda}} ,
\end{equation}
implies that $ {\cal N}_{1} \sqrt{ 2} (-\Lambda)^{- 1/4}$. 
Clearly, if the brane has positive tension the gauge-invariant vector zero 
mode is never  localized. 
Consider now the evolution equation for the gauge fluctuations. 
In the presence of the dilaton field the kinetic term of the 
gauge fields picks up a coupling to the 
higher-dimensional dilaton field. 
In particular the action will be written as 
\begin{equation}
S = -\frac{1}{4} \int d^{5} x \sqrt{|G|} e^{- \varphi} F_{A B} F^{ A B}.
\label{gaugeac2}
\end{equation}
The evolution equation for the gauge modes has the form anticipated 
in the introduction namely \footnote{ Notice that the energy-momentum 
 tensor derived from the gauge kinetic term of Eq. (\ref{gaugeac}) 
has been neglected in the solution of the background equations. 
This is correct, in the present case, since the background gauge field 
vanishes. However, there are brane models where the background gauge field 
does not vanish \cite{seif}.In this case the full Einstein-Maxwell 
system should 
be consistently studied and perturbed. Furthermore, it would be 
interesting to study the possible effect of back-reaction of the 
zero mode on the obtained solution. Also in this case 
the full Einstein-Maxwell system should be analyzed.
This interesting aspect is left 
for future work.}
\begin{equation}
{\cal A}_{\mu}'' - \frac{\Delta_{2}''}{\Delta_{2}} {\cal A}_{\mu} -
 \partial_{\alpha} \partial^{\alpha} {\cal A}_{\mu} =0,
\end{equation}
where, now, 
\begin{equation}
\Delta_{2}(w) = e^{ - \frac{ \varphi}{2}} \sqrt{ a}.
\end{equation}
Consider, first, the case where the dilaton is a {\em decreasing} function of
the modulus of the bulk coordinate. Then, in this case, the gauge zero-mode 
is 
\begin{equation}
{\cal A}_{\mu}^0 = {\cal N}_2 \Delta(w) = e^{\gamma_{<} |w|},
\end{equation} 
where $ \gamma_{<}=  \frac{1}{2} ( B_{+} - A^{<})$. 
From the obtained solutions 
we can immediately say that $\gamma_{<}>0$. Hence, in this case, 
the zero mode is 
not localized. 

Consider then the situation where the dilaton is an {\em increasing} function 
of the bulk coordinate. 
In this case the gauge zero mode is given by
\begin{equation}
 {\cal A}_{\mu}^0 =  {\cal N}_3 \Delta(w) = e^{-\gamma_{>} |w|},
\label{sol1}
\end{equation}
where, now, 
\begin{equation}
\gamma_{>} = \frac{1}{2}(  A_{>} - B_{+})= \biggl(\frac{ \sqrt{6}-1}{6}\biggr)
 \sqrt{ - \Lambda}.
\label{sol2}
\end{equation}
From Eqs. (\ref{sol1})--(\ref{sol2}) the gauge zero mode is 
clearly normalizable [i.e. $\gamma_{>} >0$] since $A_{>} > B_{+}$.
The normalization constant turns out to be 
\begin{equation}
{\cal N}_{3}^2  = \frac{ \sqrt{6} -1}{6} \sqrt{-\Lambda}.
\end{equation}
The zero mode of Eq. (\ref{sol1}) is  separated from the other states 
of the continuum by a mass gap. This aspect can be understood by writing down 
the equation for the mass eigenstates, namely
\begin{equation}
\biggl(- \frac{d^2 }{dw^2} + V(w) \biggr) {\cal A}_{\mu} = m^2 {\cal A}_{\mu},
\end{equation}
where $V(w)= \Delta''/\Delta \equiv \gamma_{>}^2 - 2 \gamma_{>} \delta(w)$. 
The 
massive wavefunctions start for $m^2 > \gamma_{>}^2$.

\renewcommand{\theequation}{5.\arabic{equation}}
\setcounter{equation}{0}
\section{Concluding remarks}

In this paper the localization properties of the gauge  modes coming from the 
geometry has been analyzed under the hypothesis 
that the four-dimensional Planck mass is not finite. It has been shown that
the spin one fluctuations can be localized on negative tension branes.
The gauge-invariant vector modes of the geometry are always massless.
Furthermore the localization of spin one  fluctuatuations coming from 
the gauge sector has been analyzed on the same 
type of configurations. It has been shown that solutions describing 
negative tension branes can simultaneously localize the gauge fields 
and the vector modes of the geometry. 

\section*{Acknowledgments}

It is a pleasure to thank M. E. Shaposhnikov for important 
hints which motivated the present analysis.

\end{document}